\documentclass[prl,twocolumn,superscriptaddress]{revtex4-1}
\pdfoutput=1
\usepackage{graphicx}
\graphicspath{ {} }
\usepackage{amsmath,graphicx}
\usepackage{epsfig}
\usepackage{amsmath,amssymb,mathrsfs,esint}
\usepackage[bookmarks=true,
   colorlinks=true,
   linkcolor=blue,
   urlcolor=blue,
   citecolor=blue,
   bookmarks=true,
   hyperindex=true
]{hyperref}
\usepackage{natbib}
\usepackage{relsize}
\usepackage{float}
\usepackage{dsfont}

\newcommand{\ua}{\uparrow}
\newcommand{\da}{\downarrow}
\newcommand{\sign}{\mathrm{sign}}
\newcommand{\md}[2][]{N_{#1}\!\left(#2\right)}   
\newcommand{\ket}[1]{| #1\rangle}
\newcommand{\braket}[1]{\langle #1 \rangle}
\def\be{\begin{equation}}
\def\ee{\end{equation}}

\begin{document}

\title{Itinerant ferromagnetism in 1D two-component Fermi gases}

\author{Yuzhu Jiang}
\affiliation{State Key Laboratory of Magnetic Resonance and Atomic and Molecular Physics, Wuhan Institute of Physics and Mathematics, Chinese Academy of Sciences, Wuhan 430071, China}

\author{D.V. Kurlov}
\affiliation{Van der Waals-Zeeman Institute, Institute of Physics, University of Amsterdam,
Science Park 904, 1098 XH Amsterdam, The Netherlands}

\author{Xi-Wen Guan}
\affiliation{State Key Laboratory of Magnetic Resonance and Atomic and Molecular Physics, Wuhan Institute of Physics and Mathematics, Chinese Academy of Sciences, Wuhan 430071, China}
\affiliation{Department of Theoretical Physics, Research School of Physics and Engineering,
Australian National University, Canberra ACT 0200, Australia}

\author{F. Schreck}
\affiliation{Van der Waals-Zeeman Institute, Institute of Physics, University of Amsterdam,
 Science Park 904, 1098 XH Amsterdam, The Netherlands}

\author{G.V. Shlyapnikov}
\affiliation{State Key Laboratory of Magnetic Resonance and Atomic and Molecular Physics, Wuhan Institute of Physics and Mathematics, Chinese Academy of Sciences, Wuhan 430071, China}
\affiliation{Van der Waals-Zeeman Institute, Institute of Physics, University of Amsterdam,
 Science Park 904, 1098 XH Amsterdam, The Netherlands}
\affiliation{LPTMS, CNRS, Univ. Paris-Sud, Universit\'e Paris-Saclay, 91405 Orsay, France}
\affiliation{Russian Quantum Center, Novaya Street 100, Skolkovo, Moscow Region 143025, Russia}


\begin{abstract}
We study a one-dimensional two-component atomic Fermi gas with an infinite intercomponent contact repulsion. It is found that adding an attractive resonant odd-wave interaction breaking the rotational symmetry one can make the ground state ferromagnetic. A promising system for the observation of this itinerant ferromagnetic state is a 1D gas of $^{40}$K atoms, where 3D $s$-wave and $p$-wave Feshbach resonances are very close to each other and the 1D confinement significantly reduces the inelastic decay.
\end{abstract}


\maketitle

Itinerant ferromagnetism of degenerate spin-1/2 fermions is an intriguing problem promoting our understanding of strongly correlated systems \cite{Lohneysen2007}.  
The origin of such ferromagnetic states is deeply rooted in quantum mechanics. In contrast to ultracold bosons, degenerate fermions try to avoid the ferromagnetic state because it requires them to have a significantly higher kinetic energy than in non-ferromagnetic states. Ultracold gases of atomic fermions which are in two internal states can be mapped onto spin-1/2 fermions treating the internal energy levels as pseudo-spin states. The ferromagnetic phase is the one where all atoms are in the same superposition of the two internal states, and one has a system of identical fermions. The kinetic energy is then higher than, for example, in the paramagnetic phase representing a statistical mixture of the two spin components.    

Itinerant ferromagnetism for fermions is studied since the 1930-ths, when the Stoner criterion for ferromagnetism in a free electron gas was introduced \cite{Stoner1933, Mattis1965}. In three dimensions the ground state can be ferromagnetic if there is a strong intercomponent repulsion in the paramagnetic state, which compensates the large difference in the kinetic energies of the ferro- and paramagnetic states. For atomic Fermi gases the Stoner mechanism was discussed in a number of papers~\cite{StonerMechanism} and was then tested by Monte Carlo calculations \cite{Pilati2010, Chang2011} which found an instability on approach to the strongly interacting regime. The efforts to stabilize the ferromagnetic state experimentally did not succeed \cite{Jo2009, Sanner2012}. The reason is \cite{Jo2009, Sanner2012, Pekker2011} that in three dimensions a large intercomponent repulsion corresponds to a very large and positive $s$-wave scattering length, and there is a weakly bound dimer of two fermions belonging to different internal states. A composition of such dimers has a lower energy than the ferromagnetic state even if the dimers overlap with each other, and at common gas densities the formation of the dimerized phase out of the cloud of atoms is very fast.

In one and two dimensions the difference in the kinetic energies of the ferro- and non-ferromagnetic states is even larger than in 3D. For example, in the 1D two-component Fermi gas even at an infinite intercomponent contact repulsion the energies of these states are equal, and at any finite repulsion the antiferro- or paramagnetic states have a lower energy \cite{note2}. Therefore, it looks such that in low dimensions making the ground state ferromagnetic is harder than in 3D.     
However, we show that in 1D the odd-wave interaction (analog of $p$-wave in higher dimensions) can drastically change the situation and make the ground state ferromagnetic. This interaction is momentum dependent and does not fall into the class of interactions satisfying the conditions of the Lieb-Mattis theorem \cite{Lieb1962} which states that in the 1D (two-component) Fermi system the ground state can not be ferromagnetic. 

To make the odd-wave interaction significant one needs a Feshbach and/or confinement-induced odd-wave resonance. At the same time, another resonance is needed in order to make the even-wave contact interaction strongly repulsive. Here we have a ``present from nature''. In the case of $^{40}$K atoms the $s$-wave resonance for the interaction between $9/2,-7/2$ and $9/2,-9/2$ states occurs at the magnetic field of 202.1 Gauss, and is very close to the $p$-wave resonance for the interaction between two $9/2,-7/2$ atoms at 198.8 Gauss \cite{Jin2002, Regal2003a}. In the fields between 198.8 Gauss and 202.1 Gauss the $s$-wave interaction is repulsive and the $p$-wave interaction is attractive. Moreover, the reduction of dimensionality to 1D decreases the inelastic decay even not far from the resonances, which is very promising for achieving itinerant ferromagnetism in the 1D gas of $^{40}$K.  

The even-wave scattering of identical fermions requires that the spatial wavefunction of colliding atoms is symmetric, and the spinor part is antisymmetric, i.e. it occurs through the singlet channel. On the contrary, in the odd-wave scattering the spatial part of the fermionic wavefunction is antisymmetric, and the spinor part is symmetric, which corresponds to the triplet state. If the odd-wave interaction is the same in all triplet states, and if it is relatively weak (such that in the expression for the corresponding interaction amplitude we can keep only the lowest term, which is proportional to~$k^2$, with~$k$ being the relative momentum of interacting atoms), while the even-wave repulsion is infinitely strong, then the problem is exactly solvable and can be mapped onto two-component bosons with $SU(2)$ spin rotation symmetry. For the latter case the ground state is known to be ferromagnetic~\cite{Li2003, Batchelor2003, Guan2007}. However, the spin rotation symmetry breaks if the odd-wave interaction is resonant, since it then depends on the spin projections of colliding particles. Therefore, in this regime the exact solution is no longer available, and in order to make conclusions about the character of the ground state we have to employ  many-body perturbation theory.


We consider a 1D two-component Fermi gas in free space and assume that the intercomponent contact (even-wave) interaction is infinitely repulsive. Since it takes place between two particles with zero total (pseudo)spin, it is only present in the non-ferromagnetic phases. Therefore, omitting the odd-wave interaction, the ferromagnetic phase represents an ideal single-component Fermi gas, with the Fermi momentum $k_F=\pi n$, and the total energy is equal to the kinetic energy:
\be     \label{Ekin}
E_f=E_{kin}=\frac{\pi^2\hbar^2n^2}{6m}N =\frac{1}{3}E_F N,
\ee
where~$n$~is the 1D density, $N$~is the total number of particles, $m$~is the mass of a particle, and $E_F=\hbar^2k_F^2/2m$ is the Fermi energy. The non-ferromagnetic phases in this case are described by the exactly solvable Yang-Gaudin model \cite{Gaudin1967, Yang1967}, and for any finite contact repulsion they have a lower energy than the ferromagnetic phase. If the repulsion is infinite, then the wavefunction vanishes when two particles approach each other at zero distance, and all spin configurations are degenerate with the energy equal to~$E_{kin}$ \cite{Gaudin1967, Yang1967, Guan2013}.

The odd-wave interaction can be both inter- and intracomponent. However, we are interested in the regime where this interaction is resonant. Since for the most important case of~$^{40}$K atoms the resonance in the odd-wave channel is present only between two atoms in the $9/2,-7/2$ states we confine ourselves to the odd-wave interaction between these states. Below the state $9/2,-7/2$ is denoted as spin-$\ua$, and the state~$9/2,-9/2$ as spin-$\da$. Moreover, we assume that although the odd-wave interaction is resonant, it is not too strong (a more precise condition will be given later), and still can be treated as perturbation. The fact that one can use a perturbative approach in a 1D odd-wave interacting system (in contrast to the even-wave interaction) finds its origin in the absence of a weakly bound state in a sufficiently shallow attractive potential. For the ferromagnetic many-body system our perturbative results perfectly agree with the existing Bethe Ansatz solution~\cite{Imambekov2010}.

Thus, to zero order the kinetic energy~$E_{kin}$ is the same in any spin configuration (as a consequence of the infinite even repulsion) and gives the main contribution to the total energy~$E$ of the system, while the odd-wave interaction provides a small correction, which we derive up to the second order in perturbation theory. For the non-ferromagnetic phases we employ the single-component momentum distribution functions~$\md[\ua]{k}$ and~$\md[\da]{k}$ that we obtain by solving numerically the Bethe Ansatz equations for the Yang-Gaudin model at an infinite intercomponent repulsion (see Supplemental Material). Considering equally populated $\uparrow$ and $\downarrow$ internal states,  $\md[\ua]{k}=\md[\da]{k}=\md{k}$, we have
\be  \label{norm}
\int_{-\infty}^{+\infty} \! \frac{dk}{2\pi} N(k)=\frac{n}{2};
\ee
\be \label{Ekin_general}
\int_{-\infty}^{+\infty} \! \frac{L\,dk}{2\pi} \frac{\hbar^2k^2}{2m} N(k) = E_{kin}/2 = \frac{\pi^2 \hbar^2 n^2}{12m}{N},	
\ee
with $L$ being the size of the system. For the ferromagnetic phase we use the Fermi step momentum distribution~$\md{k} = \theta\left(k_F-|k|\right)/2$. The kinetic energy obtained by using the calculated non-ferromagnetic distributions directly in Eq.~(\ref{Ekin_general}) differs from $E_{kin}$ by less than $0.3 \%$ in the antiferromagnetic phase and by approximately $0.5 \%$ in the paramagnetic phase.

\begin{figure}[h]
\centering
\includegraphics[trim = 0cm 0cm 0cm 0cm, clip=true, width = 0.95\linewidth]{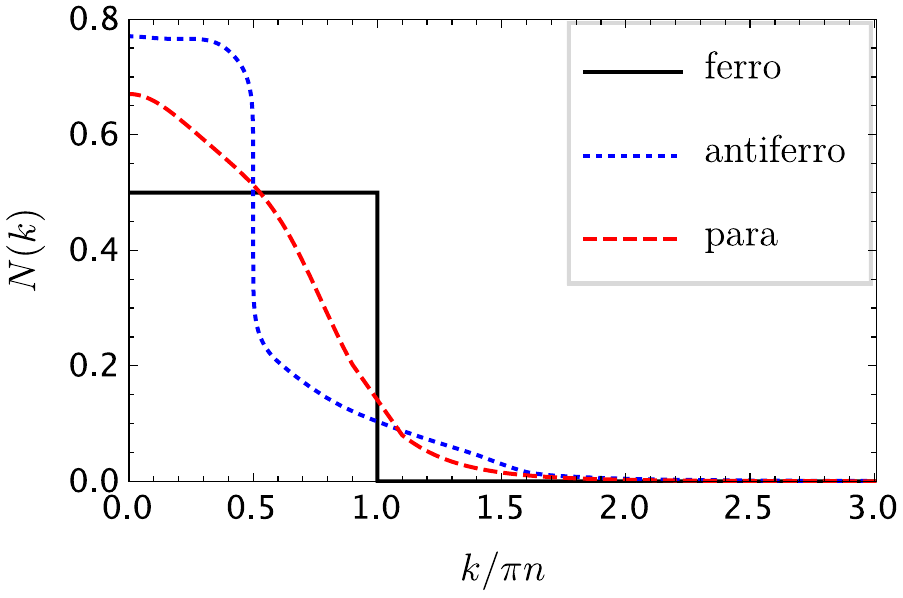}
\caption{Momentum distributions in the ferro-, antiferro-, and paramagnetic phases. The latter two are for the Yang-Gaudin model at an infinite repulsion.}
\label{momdistrfig}
\end{figure}

In order to develop many-body perturbation theory we follow the method used in Refs.~\cite{Abrikosov1958, Lu2012}. We define the off-shell scattering amplitude
\be \label{scatamp_general}
f(k',k) = \int_{-\infty}^{\infty}\!\!\!dx\, e^{-i k'\!x} \, V(x) \, \psi_k (x),
\ee
where $V(x)$ is the interaction potential, $\psi_k (x)$ is the true wavefunction of the relative motion with momentum~$k=(k_1-k_2)/2$, with $k_1$, $k_2$ and $k'_1$, $k'_2$ being the particle momenta in the incoming and outgoing scattering channels. For~$|k'| = |k'_1 - k'_2|/2 = |k|$ we have the on-shell amplitude.
Then, the total energy is $E=E_{kin}+\tilde{E}^{(1)} + \tilde{E}^{(2)}$, where the first- and second-order corrections are given by (see Supplemental Material):
\be \label{E1general}
    \mkern-22mu \tilde{E}^{(1)} = \frac{1}{L}\mathlarger{\sum_{k_1,k_2}}\tilde{f}_{odd}(k) N(k_1)N(k_2),
\ee
\be \label{E2general}
\begin{aligned}
    \mkern-10mu \tilde{E}^{(2)} &= -\frac{1}{L^2}\mathlarger{\mathlarger{\sum_{k_1,k_2,k'_1}} }\frac{4m}{\hbar^2} \frac{\tilde{f}_{odd}(k',k)\tilde{f}_{odd}(k,k')}{k_1^2 + k_2^2 - k^{'2}_1 - k^{'2}_2}\\
    &\times N(k_1)\,N(k_2)\,N(k'_1),
\end{aligned}
\ee
with $k_1+k_2 = k'_1+k'_2$. The amplitude~$\tilde{f}_{odd}$ is different from the odd-wave part of~(\ref{scatamp_general}) by the absence of the imaginary term in the denominator (see Supplemental Material). The terms~$\tilde{E}^{(1)}$ and~$\tilde{E}^{(2)}$ are the two-body (mean-field) and the many-body, or beyond mean-field, contributions to the interaction energy. As the 1D regime is obtained by tightly confining the motion of particles in two directions to zero point oscillations, the odd-wave off-shell scattering amplitude in the vicinity of the resonance is given by (see~\cite{Pricoupenko2008} and Supplemental Material):
\be \label{scatamp}
\tilde{f}_{odd}(k',k) = \frac{2\hbar^2}{m}\frac{k' k\, l_p}{1+\xi_p l_p\, k^2},
\ee
where the parameters~$l_p$ and~$\xi_p$ of the 1D odd-wave scattering can be expressed through the parameters of the 3D $p$-wave scattering as
\be  \label{1D3D}
l_p = 3a_{\perp} \left[ \frac{ a_{\perp}^3 }{w_1} + \mathcal{A} \right]^{-1}, \quad \xi_p = \frac{ \alpha_1 \, a_{\perp}^2 }{3}.
\ee
Here $w_1$ and $\alpha_1$ are the 3D scattering volume and effective range, respectively, $a_{\perp} = \sqrt{\hbar/(m \omega_{\perp})}$ is the extension of the wavefunction in the directions tightly (harmonically) confined with frequency $\omega_{\perp}$, and the numerical constant is $\mathcal{A} = -3\sqrt{2}\zeta(-1/2)\approx 0.88$, with $\zeta(-1/2)$ being the Riemann zeta function.
For $^{40}$K atoms near the $p$-wave Feshbach resonance the magnetic field dependence of $w_1$ and $\alpha_1$ has been measured in the JILA experiments \cite{Ticknor2004}. Near the resonance in 3D the scattering volume~$w_1$ changes from infinitely negative to infinitely positive, whereas the effective range~$\alpha_1$ remains practically constant and equal to $4\times10^6$~cm$^{-1}$. 
On the positive side of the resonance ($w_1>0$) in 3D one has the formation of rapidly decaying $p$-wave molecules \cite{Regal2003b, Chevy2005, Inada2008, Zhang2010, Fuchs2008}, and a similar phenomenon is expected in 1D. The issue of inelastic losses is discussed in more detail below, but in what follows we consider only the case of attractive odd-wave interaction, i.e.~$l_p<0$. The energy corrections~(\ref{E1general}) and~(\ref{E2general}) can be rewritten as (see Supplemental Material)
\be \label{E1}
\tilde{E}^{(1)} = -E_{kin}\left\{ \frac{1}{2\pi}\eta + \frac{3}{16\pi} \kappa\,\eta^2\, \mathcal{I}(Q) \right\},
\ee
\be \label{E2}
\tilde{E}^{(2)} = \frac{3}{4\pi^2}\eta^2 \mathcal{J}(Q) E_{kin} ,
\ee
where $\eta = k_F |l_p|$, $\kappa = k_F \xi_p$, $Q = \eta \kappa$, and we took into account that in any spin configuration the momentum distribution is a universal function of~$k/k_F$. The integrals~$\mathcal{I}(Q)$ and~$\mathcal{J}(Q)$ are given by
\be \label{I}
\mkern-10mu\mathcal{I}(Q) = \fint_{-\infty}^{+\infty}\! dx_1dx_2N(x_1)N(x_2)\frac{(x_1-x_2)^4}{1-\frac{Q}{4}(x_1-x_2)^2},
\ee
\be \label{J}
\begin{aligned}
\mkern-18mu \mathcal{J}(Q) &= \fint_{-\infty}^{+\infty}\! dx_1dx_2dx_3 \frac{N(x_1)N(x_2)N(x_3)}{x_1-x_3} \\
		&\mkern-30mu\times \frac{(x_1-x_2)^2}{\left(1-\frac{Q}{4}(x_1-x_2)^2\right)} \frac{(x_1+x_2-2x_3)^2}{\left(1-\frac{Q}{4}(x_1+x_2-2x_3)^2\right)},
\end{aligned}
\ee
with $x_i=k_i/k_F$ being a dimensionless momentum, and the symbol~$\fint$ denoting the principal value of the integral. The choice of a particular spin configuration is encoded in the momentum distribution~$\md{k/k_F}$, and from Eqs.~(\ref{E1})-(\ref{J}) it is evident that the odd-wave interaction splits the energies of different phases only if $\kappa\neq0$. The unperturbed momentum distributions for the\linebreak ferro-, antiferro-, and paramagnetic states are displayed in Fig.\ref{momdistrfig}. At $k \gg k_F$ the momentum distributions in the non-ferromagnetic states behave as~$\md{k}\to\mathcal{C}/k^4$, where $\mathcal{C}$~is Tan's contact~\cite{Tan2008, Barth2011}. In the aniferromagnetic phase its value is~$\mathcal{C}/k^4_F=2\ln(2)/3\pi^2\approx0.047$~\cite{Barth2011}, and in the paramagnetic phase we obtain~$\mathcal{C}/k^4_F\approx0.016$. 

In the limit of~$Q\to0$ we have $\mathcal{I} = 2(1+3\mathcal{D})/3$\linebreak and $\mathcal{J} = 1/2$, where~$\mathcal{D} = \int_{-\infty}^{\infty}\!dx\left[ N(x)x^4 - \mathcal{C}/k^4_F\right]$. For a finite~$Q$ we calculate the integrals~(\ref{I}) and (\ref{J}) numerically.

Realization of the 1D regime requires the Fermi energy to be much smaller than the tight confinement frequency:
\begin{equation}     \label{EFomega}
E_F=\frac{\hbar^2k_F^2}{2m}\ll\hbar\omega_{\perp}.
\end{equation}
For realistic confinement frequencies $\omega_{\perp}$ in the range from $50$ to $150$ kHz, the condition~(\ref{EFomega}) requires the Fermi momentum $k_F\lesssim 10^5$ cm$^{-1}$ (which corresponds to densities $n\lesssim 3\times 10^4$ cm$^{-1}$ and $E_F\lesssim 1$ $\mu$K). The confinement length~$a_{\perp}$ for such frequencies is from $400$ to $700$ \AA. Then, taking the potassium value $4\times 10^6$ cm$^{-1}$ for the effective range $\alpha_1$ and using relations (\ref{1D3D}) we see that the parameter $\kappa$ ranges from 1 to 5. In the perturbative regime we require~$\eta/\pi=n|l_p|\ll1$, i.e. one should not be too close to the resonance, and in order to stay within the limits of perturbation theory we put~$\eta<0.8$.

\begin{figure}[h]
\begin{minipage}{.5\textwidth}
	\centering
  	\includegraphics[trim=0cm 0cm 0cm 0cm, clip=true, width=0.85\linewidth]{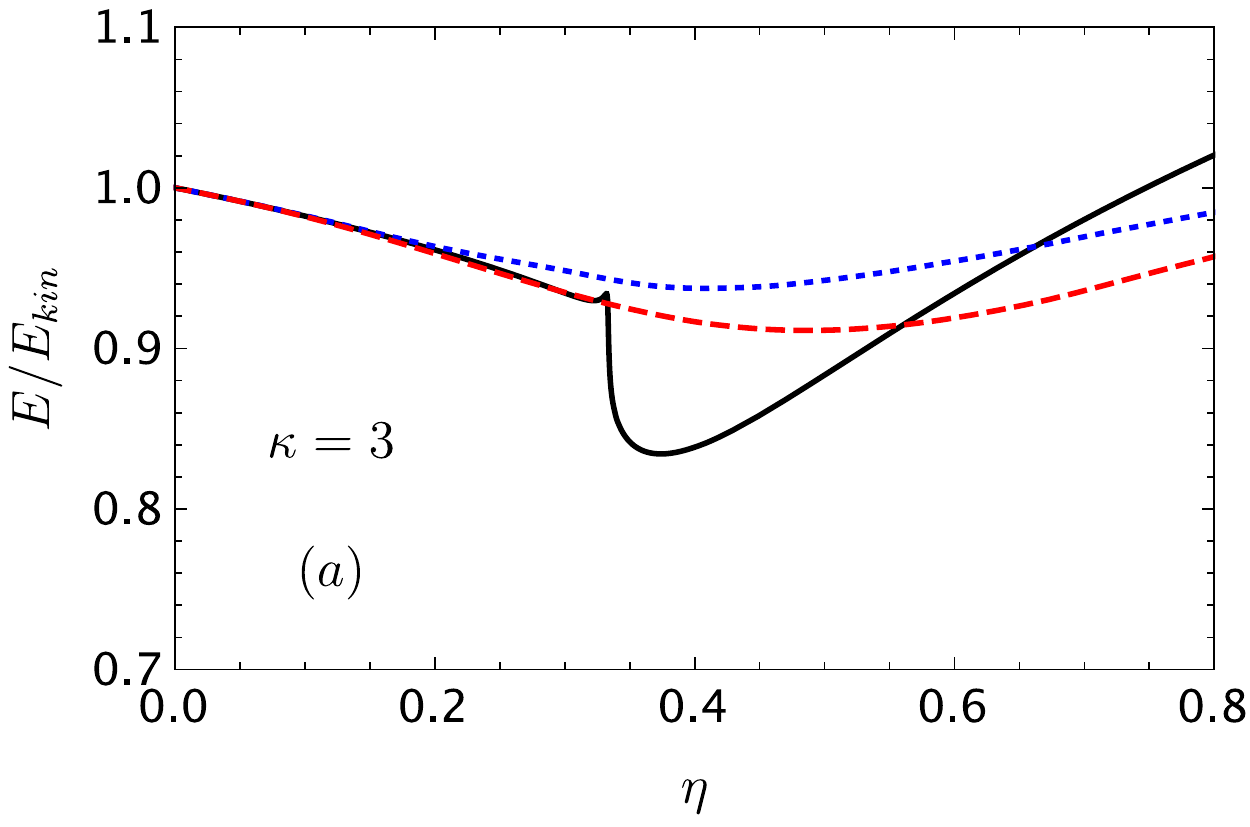}\\
\end{minipage}
\begin{minipage}{.5\textwidth}
	\centering
  	\includegraphics[trim=0cm 0cm 0cm 0cm, clip=true, width=0.85\linewidth]{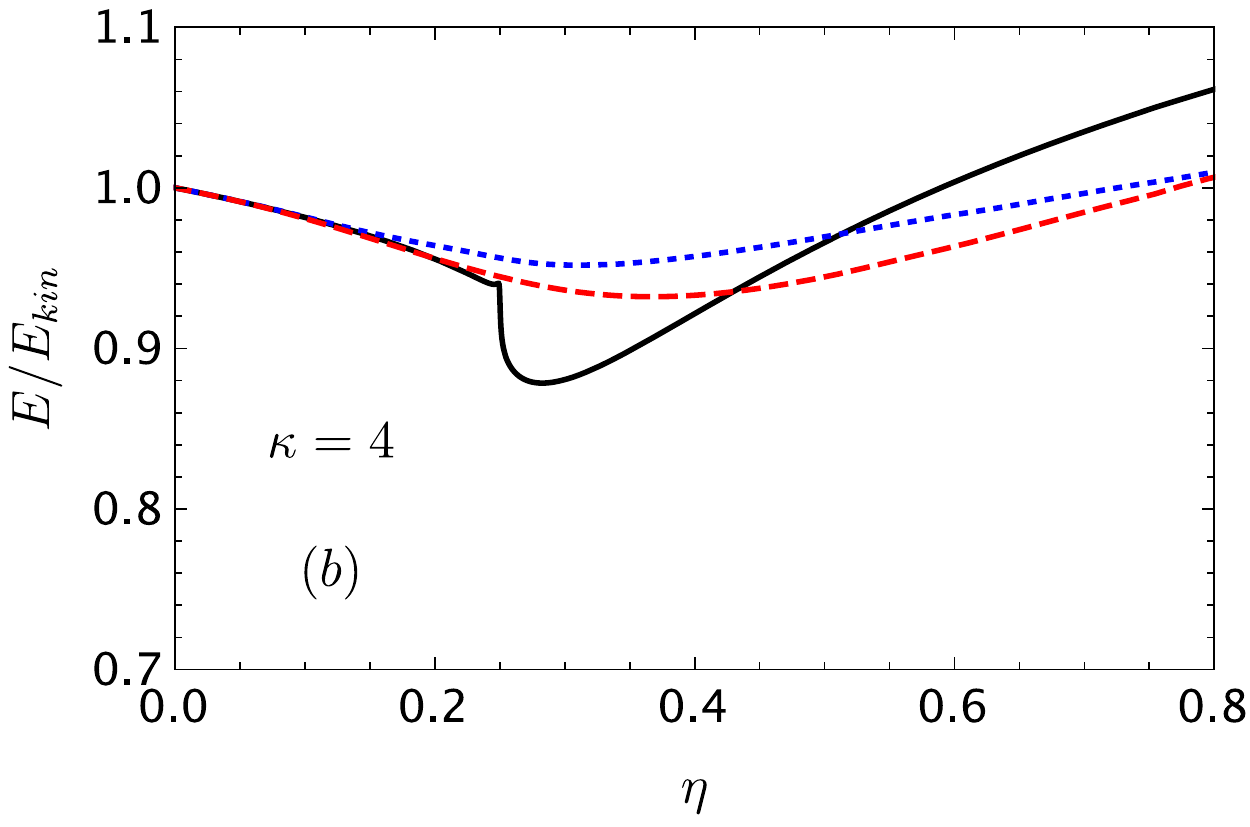}\\
\end{minipage}
\begin{minipage}{.5\textwidth}
	\centering
	\includegraphics[trim=0cm 0cm 0cm 0cm, clip=true, width=0.85\linewidth]{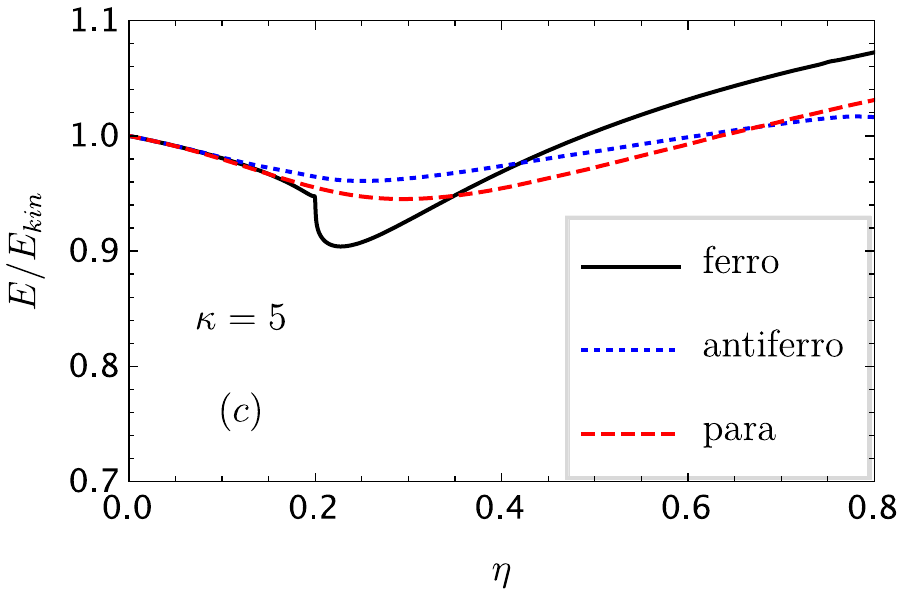}\\
	\centering
\end{minipage}
\caption{The total energy (in units of $E_{kin}=E_F N/3$) versus~$\eta$ in the ferro-, antiferro- and paramagnetic state for $\kappa=3$, $\kappa=4$, and $\kappa=5$.}
\label{EvsETA}
\end{figure}

We then calculate the total energy of the gas up to the second order in perturbation theory for the\linebreak ferro-, antiferro-, and paramagnetic phases. The results are presented in Fig.\ref{EvsETA}, which shows that the difference between the energies of the ferro- and non-ferromagnetic states is the largest at~$\eta \sim 1/\kappa$ and decreases significantly as~$\kappa$ grows. 


For a gas of~$^{40}$K atoms with a density~$n\approx3\times10^4$~cm$^{-1}$ ($E_F \approx 540$~nK) under the transverse confinement with~frequency~$\omega_{\perp} \approx 100$~kHz we have\linebreak $\kappa=\pi n \alpha_1 \hbar /(3 m \omega_{\perp})\approx3.1$, and at fields slightly lower than~$199$~G the parameter~$\eta$ is approximately~$0.36$. Then the ferromagnetic state has the lowest energy, and the energies of antiferro- and paramagnetic states are close to each other. The energy difference $(E_p - E_f)/N$ is about~$0.03 E_F$ or~$16$ nK. Increasing the confinement strength to~$\omega_{\perp} \approx 120$~kHz, we obtain~$\kappa\approx2.6$, and with~$\eta\approx0.43$ at the magnetic field $B\approx199$~G the energy difference $(E_p - E_f)/N$ becomes approximately~$20$~nK. Thus, the ferromagnetic state can be observed at temperatures below~$20$~nK.

The regimes described above ensure that the even repulsion is infinitely strong, although at first sight the corresponding magnetic fields are not close enough to the $s$-wave resonance. However, in the 1D geometry obtained by tightly confining particles in two directions, the coupling constant for the even contact interaction is \cite{Olshanii1998}
\be \label{g1D}
	g_{1D} = \frac{2\hbar^2a}{m a_{\perp}(a_{\perp}-1.03a)}.
\ee
For the confining frequencies of~$100$ and~$150$~kHz the harmonic length~$a_{\perp}$ is about~$500$ and~$400$ \AA, respectively. In a field close to 199 Gauss the scattering length\linebreak is $a\approx400$ \AA. Thus, due to the confinement-induced resonance, one can achieve an infinite contact repulsion $g_{1D}\to\infty$.

It is well known that in three dimensions $p$-wave Feshbach resonances are suffering of rapid inelastic losses~\cite{Regal2003b, Chevy2005, Inada2008, Zhang2010, Fuchs2008}.
There are two types of inelastic collisional processes. The first one is three-body recombination, which is especially pronounced if there are weakly bound dimer states. However, weakly bound dimer $p$-wave states are expected only on the positive side of the resonance ($l_p>0$), and on the negative side ($l_p<0$ and attractive interactions) the three-body recombination should not be very dangerous, at least slightly away from the resonance.  Another decay process is two-body relaxation. The internal state $9/2,-7/2$ has a higher energy than $9/2,-9/2$. Therefore, the state $9/2,-7/2$ can undergo collisional relaxation to the $9/2,-9/2$ state. In 3D the rate of this process is especially high very near the resonance \cite{Regal2003b, BohnPrivate}.

In fields slightly higher than 199 G the measured rate constant of three-body recombination in 3D\linebreak is $\alpha_{rec}^{3D}\sim10^{-25}$ cm$^6$/s and the rate constant of two-body relaxation is $\alpha_{rel}^{3D}\sim10^{-14}$ cm$^3$/s \cite{Regal2003b}.
The measurements were done in a pure gas of $9/2,-7/2$ atoms, and in a mixture of $9/2,-7/2$ and $9/2,-9/2$ states it can be somewhat higher.
The temperature in the experiment was from 1~$\mu$K to 3 $\mu$K, so that one expects about the same rate constants at $E_F\sim~1$~$\mu$K and much lower temperatures.
In order to transform these results to 1D one should recall that the inelastic processes occur at atomic interparticle distances.
We thus may integrate out the motion in the tightly confined directions~\cite{Gangardt2003}.
This leads to $\alpha_{rel}^{1D}\approx\alpha_{rel}^{3D}/2\pi a_{\perp}^2$ \cite{note}.
With the above specified $\alpha_{rel}^{3D}$ and densities $n\sim10^4$ or $3\times10^4$ cm$^{-1}$ we obtain a relaxation time of about a second.

Regarding the three-body recombination the situation is more peculiar.
For the collision of three particles in one and the same internal state in 1D one has an extra suppression by a factor of $E_F/E_*$ compared to 3D~\cite{Mehta2007}.
The quantity $E_*$ is a typical energy in the molecular problem, and one has $E_*\sim \hbar^2/mR_e^2$, where $R_e\sim50$ \AA~is the radius of interaction between particles.
So, $E_*\sim$1~mK or even larger and there is an extra suppression by 3 orders of magnitude for $E_F\sim$~1~$\mu$K.
Integrating out the particle motion in the tightly confined direction we obtain $\alpha_{rec}^{1D}\sim\alpha_{rec}^{3D}(E_F/E_*)/3\pi^2 a_{\perp}^4$ \cite{note}.
Again, for the above mentioned parameters we obtain the decay time about a second at 1D densities in between $10^4$ and $3\times10^4$ cm$^{-1}$. The ferromagnetic state can be viewed as a composition of identical fermions, and hence this estimate remains valid in this phase. 

We thus see that in 1D the issue of inelastic decay processes is not as crucial as in 3D.
In this respect the situation is somewhat similar to the one in recent experiments with strongly interacting bosons~\cite{Haller2009, Haller2010}.


In conclusion, we showed that there is a realistic possibility to find itinerant ferromagnetic states in 1D two-component Fermi gases, and a promising system is the gas of $^{40}$K atoms. This will require fine tuning of the interaction between particles by varying the magnetic field and the strength of the tight confinement. The required temperatures are about 10 to 20 nanokelvin, which is achievable with present facilities.

We would like to thank V. Gritsev, D. Petrov, and M. Zvonarev for useful discussions. We acknowledge support from IFRAF and from the Dutch Foundation FOM. The research leading to these results has received funding from the European Research Council under European Community's Seventh Framework Programme (FR7/2007-2013 Grant Agreement no. 341197).
This  work has been partially  supported by  the  NNSFC under grant numbers 11374331 and 11304357.


\onecolumngrid
\vspace{\baselineskip}
\begin{center}
\textbf{\large Supplemental Material: Itinerant ferromagnetism in 1D two-component Fermi gases}
\end{center}

\setcounter{equation}{0}
\setcounter{figure}{0}
\setcounter{table}{0}
\setcounter{page}{1}
\makeatletter
\renewcommand{\theequation}{S.\arabic{equation}}
\renewcommand{\thefigure}{S\arabic{figure}}
\renewcommand{\bibnumfmt}[1]{[S#1]}
\renewcommand{\citenumfont}[1]{S#1}

In the Supplemental Material we first discuss the scattering of identical fermions in 1D, obtain the expression for the odd-wave off-shell scattering amplitude, and derive two-body and many-body contributions to the odd-wave interaction energy of a 1D Fermi gas. We then describe our calculation of the momentum distribution in the antiferro- and paramagnetic states. 


\vspace{\baselineskip}
{\it Scattering amplitude in 1D.} The off-shell scattering amplitude, defined by Eq. (\ref{scatamp_general}) of the main text, is the sum of the even-wave and odd-wave partial amplitudes:
\be \label{S_off_shell_even_plus_odd}
	f(k',k) = f_{even}(k',k) + f_{odd}(k',k),
\ee
where
\be \label{S_f_even_odd}
	f_{even}(k',k) = \int_{-\infty}^{\infty} dx' \cos k'\!x' \,V(x')\,\psi_{even}(k, x'), \quad f_{odd}(k',k) = -i\int_{-\infty}^{\infty} dx' \sin k'\!x' \, V(x') \, \psi_{odd}(k, x'),
\ee
and $\psi_{even}$, $\psi_{odd}$ are the partial wavefunctions of the relative motion in the even-wave and the odd-wave channels, respectively. For~$|k'| = |k|$ one has the on-shell amplitudes, which follow from from Eq.~(\ref{S_f_even_odd}) putting $k' = k$. These amplitudes  enter the asymptotic expression for the wavefunction of the relative motion at interparticle separations~$|x| \to \infty$~\cite{S_LL3}:
\be \label{S_psi_asympt}
	\psi_k(x) = e^{i k x} - \frac{i m}{2 \hbar^2 k} e^{i k |x|} \bigl(f_{even}(k) + \sign(x) f_{odd}(k)\bigr).
\ee
In the quasi1D regime obtained by tightly confining the motion of particles in two directions to zero point oscillations, the on-shell scattering amplitude in the odd-wave channel was calculated in Ref.~\cite{S_Pricoupenko2008}, and it reads:
\be \label{S_f_odd}
	f_{odd}(k) = \frac{2 \hbar^2}{m} \frac{k^2}{1/l_p + ik+\xi_p k^2},
\ee
where parameters $l_p$ and $\xi_p$ are given by Eq. (\ref{1D3D}) of the main text. 

We now proceed with the derivation of the odd-wave off-shell scattering amplitude. The asymptotic form of the wavefunction~$\psi_{odd}(k,x)$ at~$|x| \to \infty$ can be written as 
\be \label{S_psi_asympt_odd}
	\psi_{odd}(k,x) \propto \sin\bigl(k|x|-\delta_{odd}(k)\bigr),
\ee
where $\delta_{odd}(k)$ is the scattering phase shift. The relation between the odd-wave phase shift and the corresponding scattering amplitude follows from Eq. (\ref{S_psi_asympt_odd}) and the odd-wave part of Eq. (\ref{S_psi_asympt}):
\be \label{S_f_to_delta}
	f_{odd}(k) = \frac{2 \hbar^2 k}{m} \frac{\tan\delta_{odd}(k)}{1+i \tan\delta_{odd}(k)},
\ee
with $\tan \delta_{odd}(k) = l_p k / \left(1 + l_p \xi_p k^2\right)$. Let us rewrite the wavefunction of the odd-wave channel $\psi_{odd}(k,x)$ in the following form:
\be
	\psi_{odd}(k,x) = \frac{i \, \tilde{\psi}_{odd}(k,x)}{1+i \tan\delta_{odd}(k)},
\ee
where $\tilde{\psi}_{odd}(k,x) = \sin kx - \sign(x) \tan\delta_{odd}(k) \cos kx$ is real. Then, the odd-wave off-shell scattering amplitude can be represented as
\be \label{S_f_to_f_tilde}
	f_{odd}(k',k) = \frac{\tilde{f}_{odd}(k',k)}{1+i \tan\delta_{odd}(k)},
\ee
where $\tilde{f}_{odd}(k',k)$ is also real:
\be \label{S_f_tilde_odd_off_shell}
	\tilde{f}_{odd}(k',k) = \int_{-\infty}^{\infty}dx' \sin k'x' V(x')\tilde{\psi}_{odd}(k,x').
\ee
 Assuming that the phase shift is small, one obtains the following relation:
\be \label{S_f_odd_approx}
	f_{odd}(k',k) = \tilde{f}_{odd}(k',k) - \frac{i m}{2\hbar^2 k}\tilde{f}^2_{odd}(k',k).
\ee
Substituting $k' = k$ into Eqs. (\ref{S_f_to_f_tilde})-(\ref{S_f_odd_approx}) we obtain similar relations for the on-shell amplitudes $f_{odd}(k)$ and $\tilde{f}_{odd}(k)$.
In the case of low-energy scattering we may put $\sin k'x' \approx k'x'$ and $\sin kx' \approx kx'$ in the expressions for $\tilde{f}_{odd}(k',k)$ and $\tilde{f}_{odd}(k)$, which shows that in the odd-wave channel one obtains the off-shell amplitude from the on-shell amplitude by simply replacing~$k$ in~(\ref{S_f_to_delta}) with~$k'$. Then, using Eqs. (\ref{S_f_to_delta}) and (\ref{S_f_to_f_tilde}) we arrive at Eq. (\ref{scatamp}) of the main text for the odd-wave off-shell scattering amplitude $\tilde{f}_{odd}(k',k)$.


\vspace{\baselineskip}
{\it Two-body and many-body contributions to the interaction energy.} An infinitely strong even-wave contact repulsion can be transferred to the boundary condition for the wavefunction. The interaction part of the Hamiltonian then contains only the odd-wave interaction, which in our case is present solely between $\uparrow$-state particles:
\be \label{S_Hint}
	\hat{\mathcal{H}}_{int} = \frac{1}{2L}\sum_{k_1,k_2,q} V(q) \, \hat{a}^{\dag}_{k_1+q}\hat{a}^{\dag}_{k_2-q}\hat{a}_{k_2}\hat{a}_{k_1},
\ee
where $\hat{a}^{\dag}_k, \hat{a}_k$ are the creation and annihilation operators of $\uparrow$-fermions, and~$V(q)$ is the Fourier transform of the interaction potential in the odd-wave channel:
\be \label{S_V(q)}
	V(q) = \int_{-\infty}^{\infty} dx \, V(x) e^{-i q x} = \int_{-\infty}^{\infty} dx \, V(x) \cos qx,
\ee
where we took into account that $V(x)$ is even. Then, the first-order correction is given by the diagonal matrix element of $\hat{\mathcal{H}}_{int}$:
\be \label{S_E1_general}
	E^{(1)} = \frac{1}{2L}\sum_{k_1,k_2} \left[ V(0) - V(k_2-k_1) \right] \md{k_1}\md{k_2}.
\ee
The second-order correction to the energy of a state $\left| j \right>$ is given by
\be
	E^{(2)}_j = \mathlarger{\sum}_{m \neq j} \frac{ \left| V_{jm} \right|^2 }{E_j - E_m},
\ee
where the summation is over the eigenstates of the non-interacting system, and the non-diagonal matrix element $V_{jm} = \left< m \right| \hat{\mathcal{H}}_{int} \left| j \right>$ is related to the scattering of two particles from the initial state~$k_1, k_2$ to an intermediate state~$k'_1, k'_2$. In our case, the symbol~$j$ corresponds to the ground state, and the symbol~$m$ to excited states.  Then, taking into account the momentum conservation law $k_1 + k_2 = k'_1 + k'_2$, we obtain the following expression for the quantity~$\left| V_{jm} \right|^2$:
\be
	\left| V_{jm} \right|^2 = \frac{1}{(2L)^2}\sum_{k_1, k_2} \left| V(k'_1 - k_1) - V(k'_2 - k_1) \right|^2 \md{k_1} \md{k_2} \bigl( 1 - \md{k'_1} \bigr) \bigl( 1 - \md{k'_2} \bigr),
\ee
and the second-order correction becomes:
\be \label{S_E2_general}
	E^{(2)} = \frac{1}{(2L)^2}\mathlarger{\sum}_{k_1, k_2, k'_1} \frac{ \left| V(k'_1 - k_1) - V(k'_2 - k_1) \right|^2 }{\hbar^2(k_1^2 + k_2^2 - k_1'^2 - k_2'^2)/2m} \md{k_1} \md{k_2} \bigl( 1 - \md{k'_1} \bigr) \bigl( 1 - \md{k'_2} \bigr).
\ee

It is evident that the second-order correction diverges at large $k'_1$ because of the term proportional to $\md{k_1}\md{k_2}$. This artificial divergence can be eliminated if one expresses $E^{(1)}$ and $E^{(2)}$ in terms of a real physical quantity --- the scattering amplitude. The relation between the Fourier component of the interaction potential and the off-shell scattering amplitude is given by~\cite{S_LL3}
\be \label{S_f_to_V}
	f(k',k) = V(k'-k) + \frac{1}{L}\mathlarger{\sum}_{k''}\frac{V(k'-k'')f(k'',k)}{E_k - E_{k''}+i0},
\ee
where $E_k$ and $E_{k''}$ are relative collision energies, and we have $E_k - E_{k''} = \hbar^2(k_1^2 + k_2^2 - k_1''^2- k_2''^{2})/2m$, with $k_1,k_2 \;(k''_1,k''_2)$ being the momenta of colliding particles in the initial (intermediate) state. Equation (\ref{S_f_to_V}) allows us to rewrite the first-order correction as
\be \label{S_E1_1}
	E^{(1)} = \frac{1}{2L} \sum_{k_1,k_2} \left[ f(k,k) - f(-k,k) \right] \md{k_1}\md{k_2} - \frac{1}{2L^2} \mathlarger{\sum}_{k_1,k_2,k'} \frac{\left[ V(k-k') - V(-k-k') \right]f(k',k)}{\hbar^2(k^2 - k'^2 + i0)/m}\md{k_1}\md{k_2}.
\ee
In the second term of Eq. (\ref{S_E1_1}) we represent $f(k',k)$ according to Eq. (\ref{S_off_shell_even_plus_odd}) and keep only the odd-wave part~$f_{odd}(k',k)$, because the quantity $\left[ V(k-k') - V(-k-k') \right]$ is odd in $k'$, and the terms containing $f_{even}(k',k)$ vanish after the integration over~$dk'$. Then, we replace the Fourier components by the exact scattering amplitudes, which gives $\left[ V(k-k') - V(-k-k') \right] = \left[ f(k,k') - f(-k,k') \right] = 2 f_{odd}(k,k')$, and the first-order correction becomes:
\be \label{S_E1_2}
	E^{(1)} = \frac{1}{L} \sum_{k_1,k_2} f_{odd}(k) \md{k_1}\md{k_2}-\frac{1}{L} \mathlarger{\sum}_{k_1,k_2} \md{k_1}\md{k_2} \int_{-\infty}^{\infty} \frac{dk'}{2\pi} \frac{m}{\hbar^2} \frac{f_{odd}(k',k)f_{odd}(k,k')}{k^2 - k'^2 + i0}.
\ee
The contribution of the pole at $k' = k$ to the integral in Eq.~(\ref{S_E1_2}) gives $- i m f^2_{odd}(k)/2\hbar^2 k$ for each term in the sum over $k_1$, $k_2$, and we can use here $\tilde{f}_{odd}(k)$ instead of $f_{odd}(k)$. At the same time, the second term of the right hand side of  Eq.~(\ref{S_f_odd_approx}), being substituted into the first term of Eq.~(\ref{S_E1_2}), cancels the contribution of the pole in the second term of Eq.~(\ref{S_E1_2}). Therefore, we use the amplitude $\tilde{f}_{odd}(k)$ in the first term of Eq.~(\ref{S_E1_2}) and take the principal value of the integral in the second term. This leads to the following expression for the first-order correction: 
\be \label{S_E1_3}
	E^{(1)} = \frac{1}{L} \sum_{k_1,k_2} \tilde{f}_{odd}(k)\md{k_1}\md{k_2} - \frac{1}{L^2} \mathlarger{\sum}_{k_1,k_2,k'_1} \frac{2m}{\hbar^2} \frac{\tilde{f}_{odd}(k',k) \tilde{f}_{odd}(k,k')}{k_1^2 + k_2^2 - k_1'^2 - k_2'^2}\md{k_1}\md{k_2}.
\ee
The second-order correction can also be expressed in terms of the scattering amplitude. Taking into account that $V(k'_1-k_1) = V(k'-k)$ and $V(k'_2-k_1) = V(-k-k')$, we write $\left|V(k'_1 - k_1) - V(k'_2 - k_1)\right|^2 = \left[ V(k-k') - V(k+k') \right] \,\, \left[ V(k'-k) - V(-k-k') \right]$, where we used the fact that Fourier components $V(q)$ are real and even functions. Then for the second-order correction we obtain:
\be \label{S_E2_2}
	E^{(2)} = \frac{1}{L^2}\mathlarger{\sum}_{k_1, k_2, k'_1} \frac{2m}{\hbar^2} \frac{ f_{odd}(k',k)f_{odd}(k,k') }{k_1^2 + k_2^2 - k_1'^2 - k_2'^2} \md{k_1} \md{k_2} \bigl( 1 - \md{k'_1} \bigr) \bigl( 1 - \md{k'_2} \bigr),
\ee
where we may use the amplitudes~$\tilde{f}_{odd}(k',k)$ and $\tilde{f}_{odd}(k,k')$ because the contribution of~$\tan \delta_{odd}(k)$ in the denominator of Eq.~(\ref{S_f_to_f_tilde}) is negligible. Then, the divergent term proportional to $\md{k_1}\md{k_2}$ in Eq.~(\ref{S_E2_2}) and the (divergent) second term of Eq.~(\ref{S_E1_3}) exactly cancel each other. Note that in Eq.~(\ref{S_E2_2}) the term proportional to the product of four occupation numbers vanishes, since its numerator is symmetrical and the denominator is antisymmetrical with respect to an interchange of $k_1, k_2$ and $k'_1, k'_2$.  Two terms containing the product of three occupation numbers are equal to each other, because the expression (\ref{S_E2_2}) for~$E^{(2)}$ is symmetrical with respect to an interchange of $k'_1$ and $k'_2$. Therefore, the sum of the first- and second-order corrections can be written as $E^{(1)} + E^{(2)} = \tilde{E}^{(1)} + \tilde{E}^{(2)}$, where $\tilde{E}^{(1)}$ and $\tilde{E}^{(2)}$ are given by 
\be \label{S_E1_4}
	\tilde{E}^{(1)} = \frac{1}{L} \sum_{k_1,k_2} \tilde{f}_{odd}(k)\md{k_1}\md{k_2} = \frac{2\hbar^2}{mL}\sum_{k_1,k_2} \frac{l_p k^2}{1+l_p \xi_p k^2} \md{k_1}\md{k_2},
\ee
\be \label{S_E2_3}
\begin{aligned}
	\tilde{E}^{(2)} &= -  \frac{1}{L^2}\mathlarger{\sum}_{k_1, k_2, k'_1} \frac{4m}{\hbar^2} \frac{ \tilde{f}_{odd}(k',k)\tilde{f}_{odd}(k,k') }{k_1^2 + k_2^2 - k_1'^2 - k_2'^2} \md{k_1} \md{k_2} \md{k'_1}\\
	&= - \frac{16\hbar^2}{m L^2} \mathlarger{\sum}_{k_1, k_2, k'_1} \frac{l^2_p \, k'^2 k^2}{k_1^2 + k_2^2 - k_1'^2 - k_2'^2}\, \frac{\md{k_1} \md{k_2} \md{k'_1}}{(1+l_p\xi_p k'^2)(1+l_p\xi_p k^2)}.
\end{aligned}
\ee
We then reduce Eqs. (S22) and (S23) to equations (9)-(12) of the main text.

\vspace{\baselineskip}
{\it Momentum distributions for the antiferro- and paramagnetic states.}
The momentum distribution functions $N_{\uparrow}(k)$ and $N_{\downarrow}(k)$ for  the antiferro- and paramagnetic states 
can be calculated from the Bethe Ansatz wave functions of the Yang-Gaudin model with an infinite repulsion. To this end, we will use the method proposed by Ogata and Shiba~\cite{S_Ogata1990}.  The  one-dimensional two-component Fermi  gas with contact interactions is described by the Hamiltonian~\cite{S_Gaudin1967, S_Yang1967}
$ \hat H =  -\sum_{i=1}^N \partial^2/\partial x_i^2
  +2c\sum_{i<j} \delta(x_i-x_j)$,
where $c=mg_{\rm 1D}/\hbar^2$ is the interaction strength for the even-wave scattering.
The Bethe Ansatz wavefunction for this model is given by \cite{S_Yang1967}
\begin{align}
 \label{psi}
 &\varPsi_{\boldsymbol{\sigma}}(\boldsymbol{x})=\sum_{{\cal Q} {\cal P}} \Theta({\cal Q}) \boldsymbol{A} ({\cal Q},{\cal P}) {\rm e}^{{\rm i} \sum_j k_{{\cal P}_j} x_{{\cal Q}_j}},
\end{align}
where ${\cal Q} =(Q1,Q2,\ldots, QN)$ and ${\cal P} =(P1,P2,\ldots, PN)$ are two  permutations of integers $\{1,2, \cdots, N\}$ and  $\Theta({\cal Q})$  denotes the  step function, i.e.  $\Theta({\cal Q}) = \theta(x_{{\cal Q}_2} - x_{{\cal Q}_1}) \theta(x_{{\cal Q}_3} - x_{{\cal Q}_2}) \cdots \theta(x_{{\cal Q}_{N}} - x_{{\cal Q}_{N-1}})$ with $\theta (x) = 1$ for  $x>0$  whereas $\theta (x) = 0$ for $x<0$.
In the above equations, we denoted 
$\boldsymbol{x}=\{x_1,x_2,\cdots,x_N\}$, $\boldsymbol{\sigma}=\{\sigma_1,\sigma_2,\cdots,\sigma_N\}$ and
$\boldsymbol{A}=A_{\sigma_1 \cdots \sigma_N} ({\cal Q},{\cal P})$ are superposition coefficients. Here  $\sigma_j=\pm 1/2$ stand for the spin projection of the $j$-th particle.

For periodic boundary conditions the wave numbers $\left\{k_j \right\}$ with $j=1,2\ldots, N$ are subject to the following  Bethe Ansatz equations (BAE):
\begin{eqnarray}
 \label{BAE}
 && {\rm e}^{{\rm i} k_jL} =
 \prod_{i=1}^M 
 \frac{k_j-\mu_{i}+{\rm i}c/2}{k_j-\mu_{i}-{\rm i}c/2},\label{BAE}\\ 
 &&
 \prod_{j=1}^N 
 \frac{\mu_i-k_{j}-{\rm i}c/2}{\mu_i-k_{j}+{\rm i}c/2}
 =-\prod_{i'=1}^M 
 \frac{\mu_i-\mu_{i'}-{\rm i}c}{\mu_i-\mu_{i'}+{\rm i}c}.\nonumber
\end{eqnarray}
Here $M$ is the number of down-spin fermions,   and $\mu_i$ with $i=1,\ldots, M$ are spin rapidities. 

We observe  that the BAE (\ref{BAE})  decouple into two parts in terms of charge and spin degrees of freedom as $c\to \infty$. The second set of equations in the BAE (\ref{BAE}) reduces to the BAE for the spin-$1/2$ Heisenberg XXX model with the scaling $\mu_\alpha  \to \mu_\alpha c$. In this  limit, the wave function (\ref{psi}) can be simplified as 
\begin{equation}
\varPsi_{\boldsymbol{\sigma}}(\boldsymbol{x})=\sum_{{\cal QP} }(-1)^{\cal Q+P} \varPhi \left(y_1,\ldots, y_M\right)  {\rm e}^{{\rm i} \sum_j k_{{\cal P}_j} x_{{\cal Q}_j}}.\label{WF2}
\end{equation}
Here $ \varPhi \left(y_1,\ldots, y_M\right) $ is the eigenstate of the spin XXX model with $M$ down-spins in the $N$-site lattice \cite{S_Yang1967, S_Ogata1990}.

Using the wave function (\ref{WF2}), we first  calculate  the momentum distribution in an analytical fashion. Then the final momentum distributions can be obtained by further numerical calculation. 
The momentum distribution is defined as 
 ${\cal N}_\sigma(k)=\braket{\hat \psi^\dag_\sigma(k) \hat \psi_\sigma(k)}$,
which  is the Fourier transform of the density matrix  $
{\cal N}_\sigma(k) = \iint{\rm d}y {\rm d}y' {\cal N}_\sigma(y', y){\rm e}^{{\rm i}k(y-y')}$.
The density matrix is given by
\begin{eqnarray}
 \label{dm-0}
  {\cal N}_\sigma(y', y) &=&\braket{\psi^\dag_\sigma(y') \psi_\sigma(y)}
 =N! \int_{0}^{x_3} {\rm d}x_2 \int_{0}^{x_4} {\rm d}x_3 \cdots \int_{0}^{x_{N}} {\rm d}x_{N-1} \int_{0}^{L} {\rm d}x_{N}
 \sum_{\sigma_2 \cdots \sigma_N}
 \varPsi_{\boldsymbol{\sigma}}^*(\boldsymbol{x}^{(l)}) 
 \varPsi_{\boldsymbol{\sigma}}  (\boldsymbol{x}^{(r)}),
\end{eqnarray}
where we used the following notations:
\begin{eqnarray}
\boldsymbol{x}^{(l)}&=&\{y',x_2,x_3,\cdots,x_N\},\,\,\, \boldsymbol{x}^{(r)}=\{y,x_2,x_3,\cdots,x_N\},\,\,\,\boldsymbol{\sigma}=\{\sigma,\sigma,\sigma_3,\cdots,\sigma_N\}.
\end{eqnarray}
In the limit $c\to\infty$,  the superposition coefficients satisfy the relations \cite{S_Ogata1990}:
\begin{equation}
 \vec{\boldsymbol A}({\cal Q}^{(ab)},{\cal P}^{(ab)})
 =\hat P_{ab}\vec{\boldsymbol A}({\cal Q}^{(ba)},{\cal P}^{(ba)}),\,\,\, \vec{\boldsymbol A}({\cal Q}^{(ab)},{\cal P}^{(ab)})
=-\vec{\boldsymbol A}({\cal Q}^{(ab)},{\cal P}^{(ba)}),\nonumber
\end{equation}
where 
\begin{eqnarray}
{\cal Q}^{(ab)}=\{{\cal Q}_1,{\cal Q}_2,\cdots,{\cal Q}_a,{\cal Q}_b,\cdots, {\cal Q}_N\},\,\,\,{\cal Q}^{(ba)}=\{{\cal Q}_1,{\cal Q}_2,\cdots,{\cal Q}_b,{\cal Q}_a,\cdots, {\cal Q}_N\}.\nonumber
\end{eqnarray}

Substituting the wave function  (\ref{WF2}) into  (\ref{dm-0}), the density matrix is thus rewritten as
\begin{eqnarray}
 \label{bethedm}
 {\cal N}_\sigma(y',y)&=&N! \int_{0}^{x_3} {\rm d}x_2 \int_{0}^{x_4} {\rm d}x_3 \cdots  \int_{0}^{L} {\rm d}x_{N}
 \sum_{{\cal P},{\cal P}'}(-1)^{{\cal P}'+{\cal P}+{\cal Q}+{\cal Q}'}
 {\rm e}^{-{\rm i}\sum_{j=1}^N k_{{\cal P}'_j} x^{(l)}_{{\cal Q}'}}
 {\rm e}^{{\rm i}\sum_{j=1}^N k_{{\cal P}_j} x^{(l)}_{{\cal Q}}}
 \nonumber\\
& &\times
 \sum_{\sigma_2,\cdots, \sigma_N}
 \big[\boldsymbol{A}^{\dag} 
 \boldsymbol{\beta}^{\dag}({\cal Q}')\big]_{\sigma,\sigma_2,\cdots, \sigma_N} 
 \big[\boldsymbol{\beta}({\cal Q}) 
 \boldsymbol{A}\big]_{\sigma,\sigma_2,\cdots, \sigma_N}.
\end{eqnarray}
Here the positions $y$ and $y'$  in the domain $x_2<\cdots<x_N$ satisfy 
two inequalities:
\begin{eqnarray}
 &&x_2<x_3<\cdots<x_\eta<y'<x_{\eta+1}<\cdots<x_N, \nonumber \\
 &&x_2<x_3<\cdots<x_\xi <y <x_{\xi+1}<\cdots<x_N. \nonumber
\end{eqnarray}
Accordingly, the permutations ${\cal Q}$ and ${\cal Q}'$ are given by 
\begin{eqnarray}
 {\cal Q}'&=&\{2,3,\cdots,\eta,1,\eta+1,\cdots,N\},\,\,\, {\cal Q}=\{2,3,\cdots,\xi,1,\xi+1,\cdots N\}.
\end{eqnarray}
The values of $\eta$ and $\xi$ are fixed by the values of $y$ and $y'$ in the set of  $\{x_j\}$ with $j=2,3,\cdots,N$. 
In equation (\ref{bethedm}) we define the operator $\boldsymbol{\beta}$ as 
\begin{eqnarray}
   \boldsymbol{\beta}({\cal Q}')&=&\hat P_{\eta 1}\hat P_{\eta-1,1}\cdots \hat P_{3,1}\hat P_{2,1},\\
   \boldsymbol{\beta}({\cal Q})&=&\hat P_{\xi 1}\hat P_{\xi-1,1}\cdots \hat P_{3,1}\hat P_{2,1}.
\end{eqnarray}
Where $\hat P_{i,j}$ is the permutation operator. 

 Defining  an identity order $\mathds{1}=\{1,2,\cdots,N\}$ for the permutation operator, we find that $\boldsymbol{\beta}({\cal Q}')\mathds{1}={\cal Q}'$
and $\boldsymbol{\beta}({\cal Q})\mathds{1}={\cal Q}$.
In Eq.~(\ref{bethedm}) the superposition coefficient $\boldsymbol{A}\equiv  \boldsymbol{A}(\mathds{1},\mathds{1})$
 is an eigenstate of the XXX Heisenberg spin chain, i.e. $H_{\rm xxx} \boldsymbol{A}=E_{\rm xxx} \boldsymbol{A}$.
It has $C_N^M$ components of spin wavefunction for each ${\cal Q}$.
Each of them is characterized by the coordinates of $M$ down-spins. 
In the following, we use the eigenstate of the  Heisenberg spin chain to evaluate the integral in Eq.~(\ref{bethedm}).
For convenience, we  denote that 
\begin{eqnarray}
&& w_{\eta,\xi}^\sigma={\boldsymbol A}^\dag \hat{\boldsymbol W}_\sigma {\boldsymbol A}=  \sum_{\sigma_2,\cdots, \sigma_N}
 \big[\boldsymbol{A}^{\dag} 
 \boldsymbol{\beta}^{\dag}({\cal Q}')\big]_{\sigma,\sigma_2,\cdots, \sigma_N} \big[\boldsymbol{\beta}({\cal Q}) 
 \boldsymbol{A}\big]_{\sigma,\sigma_2,\cdots, \sigma_N},\nonumber
\end{eqnarray} 
with 
\begin{eqnarray}
\hat{\boldsymbol W}_\sigma
& =&\hat n^\sigma_\xi \hat P_{\xi,\xi+1} \hat P_{\xi+1,\xi+2} \cdots \hat P_{\eta-1,\eta}.
\end{eqnarray}
Thus $w_{\eta,\xi}^\sigma$ can be regarded as  the expectation value of the operator $\hat{\boldsymbol{W}}_\sigma$ of the XXX spin chain.
In the above equation we also denoted the operators $\hat n^\uparrow=\hat \sigma_+\hat \sigma_-$, $\hat n^\downarrow=\hat \sigma_-\hat \sigma_+$.
By using the above notations, the density matrix is rewritten as  
\begin{eqnarray}
 {\cal N}_\sigma(y',y)&=&N! \int_{0}^{x_3} {\rm d}x_2 \int_{0}^{x_4} {\rm d}x_3 \cdots  \int_{0}^{L} {\rm d}x_{N}w^{\sigma}_{\eta,\xi}
 \sum_{{\cal P},{\cal P}'}(-1)^{{\cal P}'+{\cal P}+{\cal Q}+{\cal Q}'} {\rm e}^{-{\rm i}\sum_{j=1}^N k_{{\cal P}'_j} x^{(l)}_{{\cal Q}'}}
 {\rm e}^{{\rm i}\sum_{j=1}^N k_{{\cal P}_j} x^{(l)}_{{\cal Q}}}.\label{MD-sim}
\end{eqnarray}
Due to  the translational symmetry of the system, it is convenient to introduce $x=y-y',\,\, \tau=|\eta-\xi|$ and  $w_{\eta,\xi}^\sigma=w^\sigma_{\tau}$. Then the density matrix is simplified as
\begin{eqnarray}
 \label{dm}
&& {\cal N}_\sigma(x)=N! \, {\rm e}^{{\rm i} k_0 x} \sum_{i,j}^N (-1)^{i+j} {\rm e}^{{\rm i} \Delta k I_j x} \sum_{Q}(-1)^Q
 f_m(x) \prod_{i=1}^{n}\frac{1-{\rm e}^{-{\rm i}d_i \Delta k x}}{-{\rm i}d_i \Delta k},
\end{eqnarray}
The quantities $f_m,\, d_i$, and $\Delta k$ used  in (\ref{dm}) are explained below.

In the ground state the quasi-momenta take the following values \cite{S_Ogata1990}
\begin{align}
 & k_j=I_j\Delta k +k_0,~~\Delta k=\frac{2\pi}{L},
 \nonumber\\
 & I_j=-\frac{N-1}{2},-\frac{N-3}{2},\cdots,\frac{N-1}{2}.
\end{align}
It is worth noting that the  quantum numbers $I_j$ are  different form the quantum numbers defined in the usual Bethe Ansatz equations \cite{S_Ogata1990}.
This notation  is convenient  for our calculation.
The index $i$ ($j$) in Eq. (\ref{dm}) indicates  the  quasi-momentum of the  $i$-th ($j$-th) particle which is   separated from other $k$'s, i.e. 
\begin{eqnarray}
 \label{gquannum}
 && \{k_1,k_2,\cdots,k_{i-1},k_{i+1},\cdots,k_N\},\,\,\, \{k_1,k_2,\cdots,k_{j-1},k_{j+1},\cdots,k_N\}.
\end{eqnarray}
The corresponding quantum numbers of these two sets of  $k$'s read 
\begin{equation}
 {\cal I}=\{I_1,I_2,\cdots,I_{i-1},I_{i+1},\cdots,I_N\},\,\,\,
 {\cal J}=\{J_1,J_2,\cdots,J_{j-1},J_{j+1},\cdots,J_N\}.
\end{equation}
For fixing the first set of quantum numbers ${\cal I}$, there are $(N-1)!$ permutations $Q$ for the second set ${\cal J}$.
The summation  of $Q$ in Eq. (\ref{dm}) is carried out for $(N-1)!$ permutations. 
Each of these permutations gives a set of $d$'s 
\begin{align}
 \{d'_1,\cdots,d'_{N-1}\}={\cal I}-Q{\cal J}.
\end{align}
Assuming that there are $m$ zero elements in the set $\{d'_1,\cdots\}$ we have $n=N-1-m$ nonzero  $d$'s. 
We denote the nonzero ones as $\{d_1,d_2,\cdots,d_n\}$. 
The function $f_m(x)$ introduced in (\ref{dm}) is derived explicitly:
\begin{align}
 & f_m(x)=\sum_{\tau=0}^{N-1} w^{\sigma}_\tau(-1)^\tau \sum_{t=0}^\tau C_m^j C_n^{\tau-t} (L-x)^t(-x)^{m-t},
\end{align}
where $C^a_b$ stands for  the  combinatory. This is the key simplification of the momentum distribution (\ref{MD-sim}).

In the main text, we presented the interaction energy of the ferro- and non-ferromagnetic states, calculated through the corresponding momentum distribution functions. For the antiferromagnetic state, the momentum distribution function (\ref{dm}) can be calculated by using the corresponding ground state wave function of the XXX spin chain model, denoted as $ {\boldsymbol A}_{\rm anti}$. 
We performed our calculation from the Bethe Ansatz roots of the XXX spin chain with a finite size and finite number of down-spins.

The  paramagnetic state is more peculiar. It is a mixed state which consists of all possible spin states classified by the total spin of the system \cite{S_Eisenberg2002, S_Guan2007}. 
We denote the ferromagnetic state of the XXX spin chain as $ {\boldsymbol A}^{(m)}_{\rm ferro} \propto(\hat {\boldsymbol S}^-)^m  \ket{0}$ with  $m=0,1, \cdots N$. Here the state  $\ket{0}=\ket{\uparrow,\uparrow,\cdots,\uparrow}$ is the highest  state and the total spin operator is given by $\hat {\boldsymbol S}^-=\hat {\boldsymbol S}^-_1+\hat {\boldsymbol S}^-_2+\cdots+\hat {\boldsymbol S}^-_N$.
Each  state $ {\boldsymbol A}^{(m)}_{\rm ferro}$ gives rise to the density matrix $n_{\rm ferro}^{(m)}(x)$. Here we chose the state with zero spin projection, i.e., $S_z=0$, as the ferromagnetic state. In the main text, the momentum distribution of the ferromagnetic state refers to $n_{\rm ferro}^{(N/2)}(x)$.
For the paramagnetic state, we have to consider a mixed state that contains  all  possible spin configurations for the spin chain, i.e. using all  the eigenstates $ {\boldsymbol A}_{m}$ with   $m=1,2,\cdots 2^N$.
The corresponding density matrix  is denoted as $n_m(x)$. Then   the density matrix  for the paramagnetic state is  equally weighted as  $n_{\rm para}(x)=\sum_m n_m(x)/2^N$.



\begin{thebibliography}{99}

\bibitem{Lohneysen2007}
	H. von L\"ohneysen, A. Rosch, M. Vojta, and P. W\"olfle, \href{http://dx.doi.org/10.1103/RevModPhys.79.1015}{Rev. Mod. Phys. {\bf 79}, 1015 (2007)}.
\bibitem{Stoner1933}
	E.C. Stoner, \href{http://dx.doi.org/10.1080/14786443309462241}{Phil. Mag. {\bf 15}, 1018 (1933)}.
\bibitem{Mattis1965}
	D.C. Mattis, {\it The Theory of Magnetism} (Harper \& Row, New York, 1965).
\bibitem{StonerMechanism}
	T. Sogo and H. Yabu, \href{http://dx.doi.org/10.1103/PhysRevA.66.043611}{Phys. Rev. A {\bf 66}, 043611 (2002)}; R. A. Duine and A. H. MacDonald, \href{http://dx.doi.org/10.1103/PhysRevLett.95.230403}{Phys. Rev. Lett. {\bf 95}, 230403 (2005)}; H. Zhai, \href{http://dx.doi.org/10.1103/PhysRevA.80.051605}{Phys. Rev. A {\bf 80}, 051605 (2009)}; G.J. Conduit and B.D. Simons, \href{http://dx.doi.org/10.1103/PhysRevA.79.053606}{Phys. Rev. A {\bf 79}, 053606 (2009)}, \href{http://dx.doi.org/10.1103/PhysRevLett.103.200403}{Phys. Rev. Lett. {\bf103}, 200403 (2009)}; G.J. Conduit, A.G. Green, and B.D. Simons, \href{http://dx.doi.org/10.1103/PhysRevLett.103.207201}{Phys. Rev. Lett. {\bf 103}, 207201 (2009)}; L. J. LeBlanc, J. H. Thywissen, A. A. Burkov, and A. Paramekanti, \href{http://dx.doi.org/10.1103/PhysRevA.80.013607}{Phys. Rev. A {\bf 80}, 013607 (2009)}; I. Berdnikov, P. Coleman, and S. H. Simon, \href{http://dx.doi.org/10.1103/PhysRevB.79.224403}{Phys. Rev. B {\bf 79}, 224403 (2009)}.
\bibitem{Pilati2010}
	S. Pilati, G. Bertaina, S. Giorgini, and M. Troyer, \href{http://dx.doi.org/10.1103/PhysRevLett.105.030405}{Phys. Rev. Lett. {\bf 105}, 030405 (2010)}.
\bibitem{Chang2011}
	S.-Y. Chang, M. Randeria, and N.Trivedi, \href{http://dx.doi.org/10.1073/pnas.1011990108}{PNAS {\bf 108}, 51 (2011)}.
\bibitem{Jo2009}
	G.-B. Jo, Y.-R. Lee, J.-H. Choi, C.A. Christensen, T.H. Kim, J.H. Thywissen, D.E. Pritchard, and W. Ketterle, \href{http://dx.doi.org/10.1126/science.1177112}{Science {\bf 325}, 1521 (2009)}.
\bibitem{Sanner2012}
	C. Sanner, E.J. Su, W. Huang, A. Keshet, J. Gillen, and W. Ketterle, \href{http://dx.doi.org/10.1103/PhysRevLett.108.240404}{Phys. Rev. Lett. {\bf 108}, 240404 (2012)}.
\bibitem{Pekker2011}
	D. Pekker, M. Babadi, R. Sensarma, N. Zinner, L. Pollet, M.W. Zwierler, and E. Demler, \href{http://dx.doi.org/10.1103/PhysRevLett.106.050402}{Phys. Rev. Lett. {\bf 106}, 050402 (2011)}.
\bibitem{note2}
	Proposals to obtain the ferromagnetic and spin segregated states by sweeping across a Feshbach resonance from strongly repulsive to attractive interaction in a two-component harmonically trapped 1D Fermi gas were made in S.E. Gharashi and D. Blume, \href{http://dx.doi.org/10.1103/PhysRevLett.111.045302}{Phys. Rev. Lett. {\bf 11}, 045302 (2013)}; X. Cui and T.-L. Ho, \href{http://dx.doi.org/10.1103/PhysRevA.89.023611}{Phys. Rev. A {\bf 89}, 023611 (2014)}.
\bibitem{Lieb1962}
	E.H. Lieb and D.C. Mattis, \href{http://dx.doi.org/10.1103/PhysRev.125.164}{Phys. Rev. {\bf 125}, 164 (1962)}.
\bibitem{Regal2003a}
	C.A. Regal, C. Ticknor, J.L. Bohn, and D.S. Jin, \href{http://dx.doi.org/10.1103/PhysRevLett.90.053201}{Phys. Rev. Lett. {\bf90}, 053201 (2003)}.
\bibitem{Jin2002}
	T. Loftus, C.A. Regal, C. Ticknor, J.L. Bohn, and D.S. Jin, \href{http://dx.doi.org/10.1103/PhysRevLett.88.173201}{Phys. Rev. Lett. {\bf88}, 173201 (2002)}.
\bibitem{Li2003}
	Y.-Q. Li, S.-J. Gu, Z.-J. Ying, and U. Eckern, \href{http://dx.doi.org/10.1209/epl/i2003-00183-2}{Europhys. Lett. {\bf61}, 368 (2003)}.
\bibitem{Batchelor2003}
	M.T. Batchelor, M. Bortz, X.-W. Guan, and N. Oelkers, \href{http://dx.doi.org/10.1088/1742-5468/2006/03/P03016}{J. Stat. Mech. {\bf2006}, P03016 (2003)}.
\bibitem{Guan2007}
	X.-W. Guan, M.T. Batchelor, and M. Takahashi, \href{http://dx.doi.org/10.1103/PhysRevA.76.043617}{Phys. Rev. A {\bf76}, 043617 (2007)}.
\bibitem{Gaudin1967}
	M. Gaudin, \href{http://dx.doi.org/10.1016/0375-9601(67)90193-4}{Phys. Lett. A {\bf 24}, 55 (1967)}.
\bibitem{Yang1967}
	C.-N.Yang, \href{http://dx.doi.org/10.1103/PhysRevLett.19.1312}{Phys. Rev. Lett. {\bf 19}, 1312 (1967)}.
\bibitem{Guan2013}
	X.-W. Guan, M.T. Batchelor, C. Lee, \href{http://dx.doi.org/10.1103/RevModPhys.85.1633}{Rev. Mod. Phys. {\bf 80}, 1633 (2013)}.
\bibitem{Imambekov2010}
	A. Imambekov, A.A. Lukyanov, L.I. Glazman, and V. Gritsev, \href{http://dx.doi.org/10.1103/PhysRevLett.104.040402}{Phys. Rev. Lett. {\bf 104}, 040402 (2010)}.
\bibitem{Abrikosov1958}
	A.A. Abrikosov, I.M. Khalatnikov, Sov. Phys. JETP {\bf 6}, 888 (1958).
\bibitem{Lu2012}
	Z.-K. Lu, G.V. Shlyapnikov, \href{http://dx.doi.org/10.1103/PhysRevA.85.023614}{Phys. Rev. A {\bf85}, 023614 (2012)}.
\bibitem{Pricoupenko2008}
	L. Pricoupenko, \href{http://dx.doi.org/10.1103/PhysRevLett.100.170404}{Phys. Rev. Lett. {\bf 100}, 170404 (2008)}.
\bibitem{Ticknor2004}
	C. Ticknor, C.A. Regal, D.S. Jin, and J.L. Bohn, \href{http://dx.doi.org/10.1103/PhysRevA.69.042712}{Phys. Rev. A {\bf 69}, 042712 (2004)}.
\bibitem{Regal2003b}
	C.A. Regal, C. Ticknor, J.L. Bohn, and D.S. Jin, \href{http://dx.doi.org/10.1103/PhysRevLett.90.053201}{Phys. Rev. Lett. {\bf90}, 053201 (2003)}.
\bibitem{Chevy2005}
	F. Chevy, E.G.M. van Kempen, T. Bourdel, J. Zhang, L. Khaykovich, M. Teichmann, L. Tarruell, S.J.J.M.F. Kokkelmans, and C. Salomon, \href{http://dx.doi.org/10.1103/PhysRevA.71.062710}{Phys. Rev. A {\bf 71}, 062710 (2005)}.
\bibitem{Inada2008}
	Y. Inada, M. Horikoshi, S. Nakajima, M. Kuwata-Gonokami, M. Ueda, and T. Mukaiyama, \href{http://dx.doi.org/10.1103/PhysRevLett.101.100401}{Phys. Rev. Lett. {\bf101}, 100401 (2008)}.
\bibitem{Zhang2010}
	P. Zhang, P. Naidon, and M. Ueda, \href{http://dx.doi.org/10.1103/PhysRevA.82.062712}{Phys. Rev. A {\bf82}, 062712 (2010)}.
\bibitem{Fuchs2008}
	J. Fuchs, C. Ticknor, P. Dyke, G. Veeravalli, E. Kuhnle, W. Rowlands, P. Hannaford, and C. J. Vale, \href{http://dx.doi.org/10.1103/PhysRevA.77.053616}{Phys. Rev. A {\bf77}, 053616 (2008)}.
\bibitem{Tan2008}
	S. Tan, \href{http://dx.doi.org/10.1016/j.aop.2008.03.004}{Ann. Phys. (N.Y.), {\bf323}, 2952 (2008)}.
\bibitem{Barth2011}
	M. Barth, W. Zwerger, \href{http://dx.doi.org/10.1016/j.aop.2011.05.010}{Ann. Phys. (N.Y.), {\bf326}, 2544 (2011)}.
\bibitem{Olshanii1998}
	M. Olshanii, \href{http://dx.doi.org/10.1103/PhysRevLett.81.938}{Phys. Rev. Lett. {\bf81}, 938 (1998)}.
\bibitem{BohnPrivate}
	J. Bohn (private communication).
\bibitem{Gangardt2003}
	D.M. Gangardt and G.V. Shlyapnikov, \href{http://dx.doi.org/10.1103/PhysRevLett.90.010401}{Phys. Rev. Lett. {\bf90}, 010401 (2003)}.
\bibitem{note} A detailed study of inelastic decay processes for our system will be given elsewhere.
\bibitem{Mehta2007}
	N.P. Mehta, B.D. Esry, and C.H. Greene, \href{http://dx.doi.org/10.1103/PhysRevA.76.022711}{Phys. Rev. A {\bf76}, 022711 (2007)}.
\bibitem{Haller2009}
	E. Haller, M. Gustavsson, M.J. Mark, J.G. Danzl, R. Hart, G. Pupillo, H.-C. N\"agerl, \href{http://dx.doi.org/10.1126/science.1175850}{Science {\bf352}, 5945 (2009)}.
\bibitem{Haller2010}
	E. Haller, M.J. Mark, R. Hart, J.G. Danzl, L. Reichs\"ollner, V. Melezhik, P. Schmelcher, and H.-C. N\"agerl, \href{http://dx.doi.org/10.1103/PhysRevLett.104.153203}{Phys. Rev. Lett. {\bf104}, 153203 (2010)}; E. Haller, M. Rabie, M. J. Mark, J. G. Danzl, R. Hart, K. Lauber, G. Pupillo, and H.-C.
N\"agerl, \href{http://dx.doi.org/10.1103/PhysRevLett.107.230404}{Phys. Rev. Lett. {\bf 107}, 230404 (2011)}.

\end{thebibliography}

\begin{thebibliography}{99}

\bibitem{S_LL3}
	L.D. Landau and E.M. Lifshitz, {\it Quantum Mechanics, Non-Relativistic Theory} (Butterworth-Heinemann, Oxford, 1999).
\bibitem{S_Pricoupenko2008}
	L. Pricoupenko, \href{http://dx.doi.org/10.1103/PhysRevLett.100.170404}{Phys. Rev. Lett. {\bf 100}, 170404 (2008)}.
\bibitem{S_Ogata1990}
  M. Ogata, H. Shiba, \href{http://dx.doi.org/10.1103/PhysRevB.41.2326}{Phys. Rev. B {\bf 41}, 2326 (1990)}.
  \bibitem{S_Gaudin1967}
	M. Gaudin, \href{http://dx.doi.org/10.1016/0375-9601(67)90193-4}{Phys. Lett. A {\bf 24}, 55 (1967)}.
\bibitem{S_Yang1967}
	C.-N.Yang, \href{http://dx.doi.org/10.1103/PhysRevLett.19.1312}{Phys. Rev. Lett. {\bf 19}, 1312 (1967)}.
\bibitem{S_Eisenberg2002}
	E.  Eisenberg and E.  H. Lieb, \href{http://dx.doi.org/10.1103/PhysRevLett.89.220403}{Phys. Rev. Lett. {\bf 89}, 220403 (2002)}.
\bibitem{S_Guan2007}
	X.-W. Guan, M.T. Batchelor, and M. Takahashi, \href{http://dx.doi.org/10.1103/PhysRevA.76.043617}{Phys. Rev. A {\bf76}, 043617 (2007)}.
	
\end{thebibliography}
\end{document}